\documentclass[12pt]{iopart}
\usepackage{graphics,graphicx}

\usepackage[export]{adjustbox}

\begin{document}

\title[Interface propagation in fiber bundles]{Interface propagation in fiber bundles: Local, mean-field and intermediate range-dependent statistics}



\author{Soumyajyoti Biswas${}^{(1)}$ and Lucas Goehring${}^{(1, 2)}$}
\address{
(1) Max Planck Institute for Dynamics and Self-Organization, Am Fassberg 17, G\"{o}ttingen-37077, Germany.\\
(2) School of Science and Technology, Nottingham Trent University, Clifton Lane, Nottingham,
NG11 8NS, UK. \\
}
\ead{soumyajyoti.biswas@ds.mpg.de, lucas.goehring@ntu.ac.uk}

\date{\today}

\begin{abstract}
\noindent  The fiber bundle model  is essentially an array of  elements
that break when sufficient load is applied on them. With a 
local loading mechanism, this can serve as a model for a one-dimensional
interface separating the broken and unbroken parts of a solid in mode-I fracture. The interface can propagate through the system
depending on the loading rate and disorder present in the failure thresholds of the fibers. In the presence 
of a quasi-static drive, the intermittent
dynamics of the interface mimic front propagation in disordered media. Such situations appear in diverse physical 
systems such as mode-I crack propagation, domain wall dynamics in magnets, charge density waves, contact line in wetting etc. We study the effect of 
the range of interaction, i.e. the neighborhood of the interface affected following a local perturbation, on the
statistics of the intermittent dynamics of the front. There exists a crossover from local to global
behavior as the range of interaction grows and a continuously varying `universality' in the intermediate range.  
This means that the interaction range is a relevant parameter of any resulting physics. This is particularly relevant
 in view of the fact that there is a scatter in the 
experimental observations of the exponents, in even idealized experiments on fracture fronts, 
 and also a possibility in changing the interaction range in real samples.
\end{abstract}


\maketitle

\section{Introduction}    
The fiber bundle model, introduced in Ref. \cite{first}, is a useful approach for modeling catastrophic failures in 
disordered solids from a microscopic point of view \cite{daniels, phoenix1}. It aims at  using the minimal
ingredients that nonetheless capture the 
universal statistical features associated with phenomena \cite{book1, book2, book3} such as:  the 
 breaking of disordered samples like wood \cite{wood}, glass \cite{glass}, polymeric foam \cite{polymer}, and
paper \cite{paper}; roughness of fracture fronts 
in peeling experiments with  PDMS samples \cite{pdms1, san10, chopin15}; and precursor events in the catastrophic collapse
of cliffs \cite{cliff}, landslides \cite{land} etc. In particular, it considers a set of elements having finite failure thresholds drawn from a probability 
distribution as a simple model for a disordered solid. Evolving this model requires 
 a rule of transferring the load between elements, following the 
breaking of one such element.
Given that the width of the distribution function from which the strength of the elements are drawn is finite, the 
universal macroscopic responses are determined by the range of this load redistribution \cite{rmp1}. The local (nearest neighbors, introduced in \cite{harlow})  
and mean-field (global load sharing \cite{rmp1}) limits are well studied but are rather idealized in view of the fact 
that any realistic stress concentration range in a solid is neither of these two extremes. Several attempts have been made to interpolate
between these two limits \cite{hidalgo02, pradhan05, stormo12, gjerden13, biswas15}. However, in the case of catastrophic failures,
the stress at which the system just breaks (critical stress) becomes non-zero only when the 
range of interaction is sufficiently large, so that
it suppresses any spatial stress concentration \cite{biswas15}. Hence, statistics that are qualitatively similar 
to experiments \cite{bonamy09}
are observed only in the mean-field limit.

The above case of a fiber bundle model with system-wide loading can be contrasted with the case of a model where a load can be applied locally \cite{biswas13}, for example at some internal point, or along an edge or interface.  In such cases fibers will break around the source of the load, and will then transfer their burden to other nearby fibers.  A dynamic interface will then develop between the growing region of failed fibers, and the unbroken fibers further from the load.  Essentially, this interface will now behave as the tip, or edge, of a growing crack.  In a simpler version of this model, which we will develop here, one can simply keep track of the fibers along this effectively one-dimensional interface, at any given time (see Fig. \ref{interface_schematic}).  The stress concentration and dynamics of this model is, therefore, constrained to an ‘activity front’ that separates the two hypothetical sides: on one side of this front all the material is broken and on the other side it is completely intact. Such confinement of activity can also be achieved with system-wide loading in a two dimensional model, and a gradient in the failure thresholds of the fibers \cite{stormo12} in the direction perpendicular to the desired front. Indeed, two regimes of front roughness were recently found in such a case \cite{gjerden13}, one consistent with the elastic line and the other coalescence dominated regime.

The activity front, or the failure interface,
 can be driven through the system quasi-statically, giving rise to intermittent dynamics. 
The phase transition associated with these intermittent dynamics is the depinning transition 
of the interface (see e.g. \cite{pdms1, zapperi98}). The states to either side of the corresponding critical point lie
in  the depinned phase, where the interface
moves with a finite steady state velocity, and the pinned phase, where the interface is stopped
by pinning centers (see e.g. \cite{san10}). This situation is widely observed in different physical systems, for
example the vortex lines in superconductors \cite{larkin79}, domain walls in magnetic systems \cite{zapperi98}, charge density 
waves \cite{fischer85}, contact line dynamics in wetting \cite{talon15} and, of course, 
in mode-I crack opening \cite{bouchaud93}, on which we shall focus.

Unlike the case of catastrophic failure, in the interface depinning model even the extreme case of a
nearest neighbor interaction will have a 
finite critical load at which the interface depins. It therefore gives a chance to study the effect of spatial fluctuations on
interface dynamics. Previous studies regarding interface propagation in fiber bundles include  simulations where
the fibers were fixed between a rigid ceiling and a soft bottom plate \cite{batrouni02}. 
 For a single point loading  and redistribution along the interface, the damaged region grew 
radially outwards on average, resulting in a 
continuously growing interface \cite{biswas13}. The interface length was proportional to the applied load, resulting in a steady value of the
load per fiber along the interface. The intermittent dynamics of the interface showed scale-free statistics.   

In our work, a systematic study of the roughness and avalanche dynamics of the fiber bundle model is made 
by varying the range of interaction along a propagating interface. In particular, we approximate this interface by a one-dimensional array of fibers,
and then consider the implications of different types of load redistribution
between these interfacial fibers, when one breaks. First, we consider the case of a uniform  
redistribution over a finite range that can vary from the nearest neighbors of the broken fibers, to
the entire interface (i.e. equivalent to a mean-field model).  This can include many 
different types of interaction that have a finite interaction range, for example those due to geometric effects 
like confinement to a thin sheet, plastic deformation, 
finite agglomerate size, etc. 
In such systems we demonstrate a crossover value of the range of interaction of the load redistribution that separates the local and mean-field behaviors.
Secondly, we consider a scale-free
interaction, where the load from a broken fiber is shared amongst all the fibers along the interface, in an
amount inversely proportional to the distance of the fibers from the broken one, raised to some power.  Among 
the scale-free interaction ranges,  the inverse-square interaction has been considered before \cite{pontuale13},
inspired by the stress concentration field around a crack in an infinite sample. The inverse square interaction, 
however, is not guaranteed for a finite sample \cite{hutch, bares14},
which is often a realistic scenario. For example in the fracture or debonding of a paint film, or other thin coating, the stress is
redistributed over a distance comparable to the film thickness \cite{xia00}. Also the range of the interaction
in a real material sample can be affected by
the correlation length of any disorder (see e.g. \cite{peng91}), the size of any aggregates or agglomerates in ceramics \cite{kendall} or the size of a plastic or process zone in a ductile metal \cite{barenblatt, dugdale}, for example. We find that the inverse square
interaction lies in the middle of a continuously changing set of critical exponents for the roughness and 
avalanche size distributions. For very slow decay of the interaction strength with distance, the model is mean-field-like, but it starts deviating 
from mean-field behavior as soon as the average range of the interaction is finite.

In the following, we first describe our model and then present the numerical results obtained from it. We give
analytical calculations for the nearest-neighbor interaction and show that this case is qualitatively
different from an elastic interface model approach in that limit. Finally, we make comparison 
with mode-I crack propagation experiments and offer some conclusions about the effect of finite range interactions 
on fracture problems.

\section{Model}
\label{sec:model} 
\begin{figure}[tbh]
\centering 
\includegraphics[height=6cm]{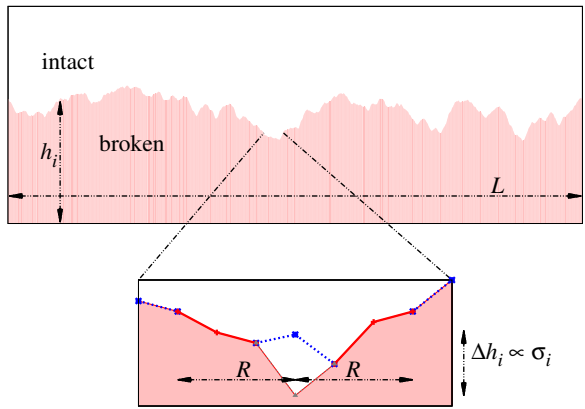}
   \caption{A snapshot of the interface between the broken (red) and unbroken (white) parts of the model system is shown. A magnified portion of 
the interface demonstrates the breaking of an element along the fracture front (dotted line) and  the subsequent advancement of the front. The load of the broken fiber is redistributed to $R=3$ neighbors (shown in red).}
\label{interface_schematic}
\end{figure}

We model the crack front of a mode-I fracture by a line of breakable fibers, as sketched in Fig. \ref{interface_schematic}. 
Essentially, we store the 
stress profile and the position of the front in a one-dimensional array, which is periodically updated.
Note that in this way we can work with a one dimensional model of a front, which is propagating on a hypothetical two dimensional plane. 
When the stress on an element exceeds its failure threshold, drawn randomly from a uniform  distribution in [$0:1$], 
there is a local failure and
the front at that point is advanced by an amount proportional to the stress at that point prior to the failure. 
Hence, while the locations of the fibers are regular and discrete along the front, they are continuous along the propagation
direction. After a fiber is advanced, the stress on it is set to zero.
 The stress that it had carried is
partly dissipated (a fraction $\delta$) and the remaining part (the fraction $1-\delta$) is redistributed along the  rest of the fracture front, as detailed below. 
Throughout this paper we have used $\delta=0.0001$ to study intermittent dynamics, and $\delta=0$ to study the effects of sudden, fixed loading. For finite $\delta$,
our results were found not to depend on the specific value of $\delta$, unless it was larger than about $0.01$.

By construction the fracture front can only move forward, hence no islands or overhangs are allowed along the front. The
total displacement of the $i$-th element, $h_i(t)$,  upto time $t$ is proportional to $\sum\limits_{t\in \mathcal{T}_f}\sigma_i(t)$,
where $\sigma_i(t)$ is the instantaneous stress at position $i$ and $\mathcal{T}_f$ is the set of  all the times where a 
fiber failed at that location. Note that the proportionality constant has no effect on the 
macroscopic measures studied here. The dynamics of the front, except for the stress 
redistribution rule, are those followed in Ref. \cite{pontuale13}.

In terms of loading our system, we will consider two loading scenarios.  First, to establish a simple point of comparison with mean-field models, a fixed load is suddenly applied to the entire system at once.  Following failure this load is then redistributed without dissipation.  In contrast, for the case of quasi-static loading, on which we focus,
 the external force increases slowly with time, like the case of peeling a piece of tape off a table.
This increase in load can cause fibers to break, and thereby advance  the fracture front.
However, in any real experiment there must be a finite dissipation of stress as the front advances. 
This can be due to acoustic emissions, friction,
heat, etc. or simply the finite compliance of any load cell. 
The dissipation will halt any advance, but the increase in load will then continue until another advance happens, 
and so on. These two competing mechanisms, loading and dissipation, lead to intermittent dynamics of the
front. In our simulations we capture this by alternating between slow loading steps and fast front activity, including dissipation. Now the load is increased until the first 
fiber breaks, then it is stopped and the load redistribution rules are followed until the avalanche of activity stops, on the assumption that internal load redistribution occurs much faster 
than the recovery of the external loading. When dissipation stops the avalanche, further loading is continued. 
In this way any system will self-organize, tending towards the critical point of 
the depinning transition in the corresponding conservative system (Fig. \ref{phase_dia}).
To avoid observing transient behavior, the first $10000$ to $100000$ avalanches (depending on the system sizes)
were excluded from analysis.

We will also explore two  types of redistribution rules, following the breaking of a fiber. 
First, we will consider the case where the redistribution only affects fibers within a finite range,
 which can be as small so as to include only the nearest neighbors, or as large as 
 the system size. For this we consider the
uniform redistribution of the load to $R$ neighbors on each side of the failed fiber, along the interface (i.e. affecting a total of $2R$ fibers).
Redistributions that affect a finite range are expected to be captured by this
process, which is also the case for the  catastrophic failure \cite{biswas15}. The two extreme cases are the nearest 
neighbor ($R=1$) and mean-field ($R\sim L$, where $L$ is the system size)
interactions and we find a crossover length scale $R_c$ above which the system starts behaving like
the mean-field limit.  

Secondly, we will consider scale-free redistribution, i.e. if the fiber at the
$i$-th site fails then a fiber at the $j$-th site will receive a load proportional to $1/|i-j|^{\gamma}$. 
Here, for small enough values of $\gamma$, and hence more long-range coupling, the interaction is expected to be mean-field. For sharp
enough decay (large $\gamma$ values) the behavior will be local. Once again, a crossover $\gamma_c=1$
is found above which the behavior of the activity front starts deviating from mean-field limit. The system also then shows a continuous variation 
in its critical exponent values for roughness and avalanche size, until it reaches the other limit, a nearest neighbor interaction,
for large enough $\gamma$. 

\section{Results} 
We report on the effects of two types and redistribution rules here, following the breaking of a fiber.  These involve either the redistribution of the load evenly over a finite range (sec. 3.1 and 3.2), or a scale-free redistribution that decays with some characteristic power-law (sec. 3.3).  For the finite-range redistribution, the case without dissipation (sec. 3.1) is considered separately from the quasi-static case (sec. 3.2).

\subsection{Depinning transition with non-dissipative finite range interaction}
\begin{figure}[tbh]
\centering 
\includegraphics[width=7.5cm, valign=t]{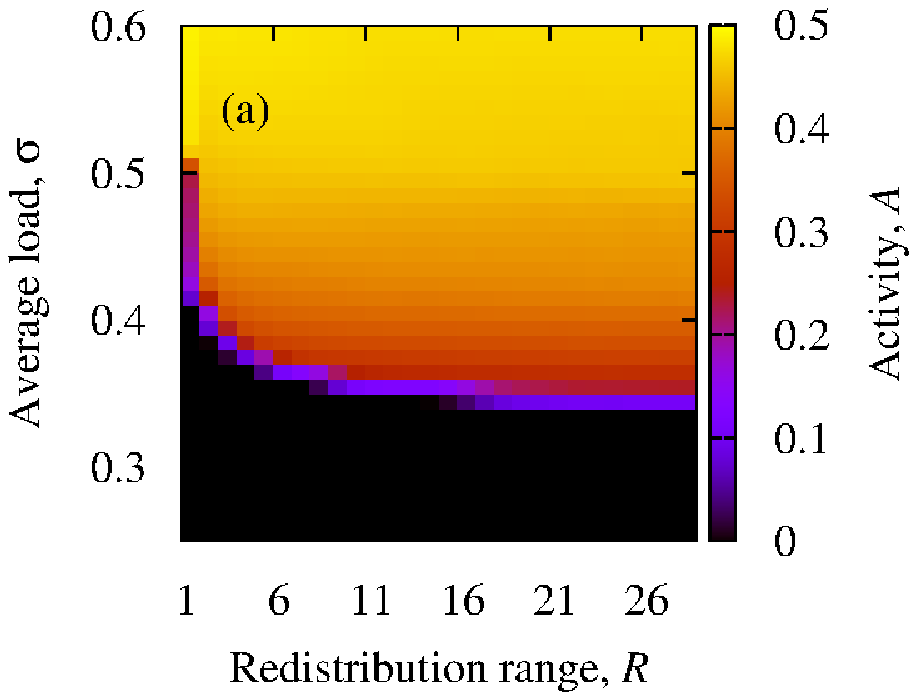}
\includegraphics[width=7.8cm, valign=t]{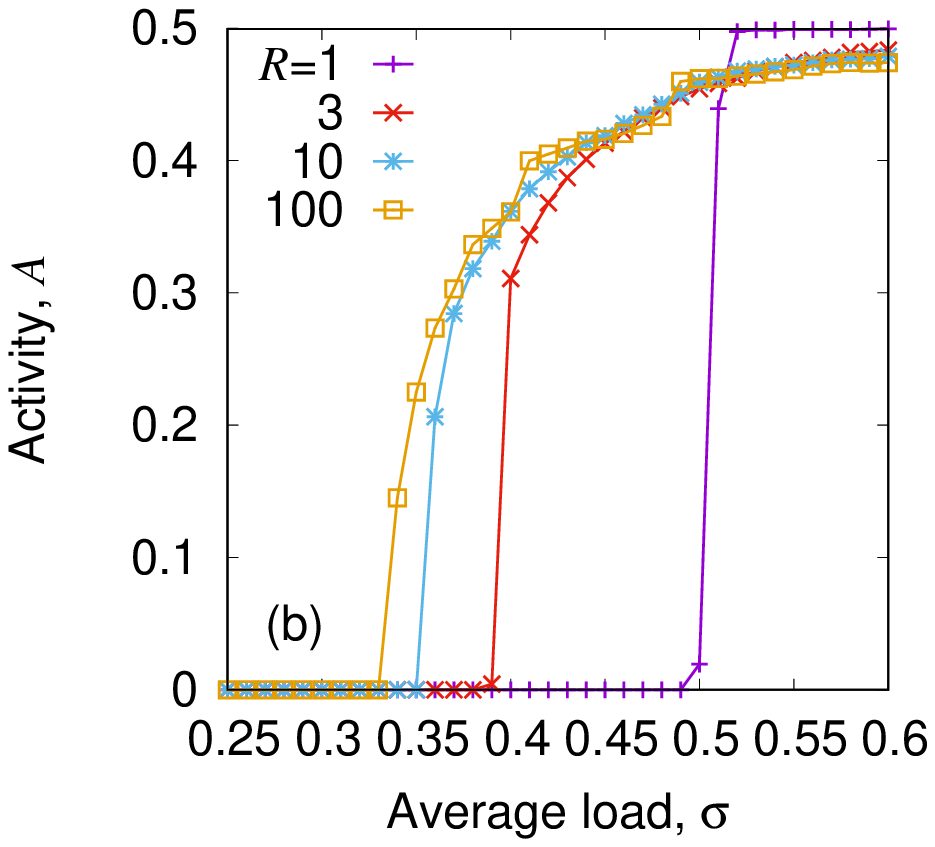}
   \caption{(a) The  phase diagram for the depinning transition with a range dependent 
load transfer. The color scale gives the steady-state breaking rate or activity, $A$ for simulations run at different loads $\sigma$ and ranges $R$. As can be seen, the nearest
neighbor case $R=1$ shows a sharp jump in the activity at the critical point $\sigma=0.5$. In the presence of
dissipation ($\delta>0$), and for a slow quasi-static drive, the critical point becomes an attractive fixed point, causing the
load per fiber to saturate at its critical value. (b) Activity rates for various values of $R$ show how its discontinuous
jump for $R=1$ gradually disappears as $R$ becomes large. The system size is $L=1000$.}
\label{phase_dia}
\end{figure}

Let us first look into the case where the load redistribution is of finite range. 
Following a failure, the load is redistribution to $2R$ nearby fibers, without dissipation. 
The fracture front propagating through the system shows a depinning transition \cite{bonamy11}
when sufficient load is applied. This depinning transition is seen in the limit of 
 no dissipation ($\delta=0$) or, in practice, when dissipation is negligibly small as compared to the total load ($\delta\ll 1$). 
It is an active-absorbing kind of phase transition \cite{henkel14}.
 The order parameter is defined here as the rate of activity, $A$,  or the number of fibers breaking per Monte Carlo 
time-step. For a large enough load it reaches a stationary non-zero value (active state), on average. When the load per 
fiber is below some critical value $\sigma_c$, it falls to zero (absorbing state).   
Fig. \ref{phase_dia}(a) shows the phase boundary between these active and absorbing states for different values of the interaction range,
$R$. The color (shade) gradient gives the activity rate $A$. The $R=1$ case shows a sharp jump [see Fig. \ref{phase_dia}(b)], 
suggesting
a discontinuous transition there.

  We will now look more carefully into the critical behavior for the nearest neighbor ($R=1$) case without dissipation. 
Consider the situation where  a load $\sigma$, per fiber, is applied uniformly and suddenly on the system. All fibers
with a failure threshold below $\sigma$ will break. This will create  broken patches, the largest of which 
is most likely to contribute to further breaking the system, leading to depinning.
If all the stress from this patch, of length $l$, is redistributed to the two neighboring points on either side 
of the patch, these points will now each carry a load $\sigma+\frac{\sigma l}{2}$. 
  Since $l$ is large, this will likely cause further 
breaking and in the next time step the increased stress at the edge of the growing active region 
 will be $\sigma+\frac{1}{2}\left(\sigma+\frac{l\sigma}{2}\right)$.
After $m$ such steps, the load on the sides of the patch will be
\begin{eqnarray}
\sigma_E &=& \left(\sigma+\frac{\sigma}{2}+\frac{\sigma}{4}+\dots+\frac{\sigma}{2^m}\right)+\frac{l\sigma}{2^m} \nonumber \\
&=& 2\sigma+\frac{\sigma(l-1)}{2^m}.
\end{eqnarray}
The second term is an exponential decay, which vanishes for large $m$. The first term will cause depinning 
i.e. breaking all possible pinning centers, only when the stress
exceeds the maximum pinning threshold 1, giving a critical load $\sigma_c=1/2$. 
The above calculation is valid in the sub-critical limit, where large $m$ values are possible and therefore
gives the limiting stress that the system can withstand without depinning.
This is the critical load that
 we see in the simulations as well (see e.g. Fig. \ref{phase_dia} for $R=1$). 
Also, at the critical load $1/2$, it is trivially seen that half of the system on average will break in the
first  time step following the loading. After redistribution of this load, the surviving fibers will have stress $1$ on average and will then
 break in the following time step. These two steps will continue to repeat in the steady 
state, giving an average activity rate of
$1/2$. Note that this is a discontinuous jump from zero activity in the absorbing phase (for $\sigma<\sigma_c$). This is in contrast to
nearest neighbor elastic line (Edwards-Wilkinson model \cite{edwards82}) depinning, which shows a continuous transition \cite{amaral95} instead.

The other extreme case, for a response to a sudden loading,  is the continuous depinning transition in the mean-field limit ($R\sim L$). Unsurprisingly, 
the mean-field exponents are recovered
there (see Appendix A).

\subsection{Intermittent dynamics with dissipative finite range interaction}
Let us now explore the critical response to  quasi-static loading in the presence of dissipation. 
After an initial failure 
 the load is redistributed, as described in sec.\ref{sec:model}, and a fraction 
of the load from each failing fiber is dissipated.
The dissipation implies that the front propagation will eventually halt, at which point we resume 
increasing the load, slowly, to restart the dynamics.

We will study the dynamic (avalanche statistics) and static (roughness of the front) properties  
of the dissipative system for different load redistribution rules, including the local and mean-field
limits, and intermediate conditions. We aim to understand the crossover dynamics between the two very 
different end-members of this model, and to determine the redistribution range at which the model behavior shifts from 
local to mean-field.

\subsubsection{Avalanche statistics:} 
\begin{figure}[tbh]
\centering 
\includegraphics[height=6cm]{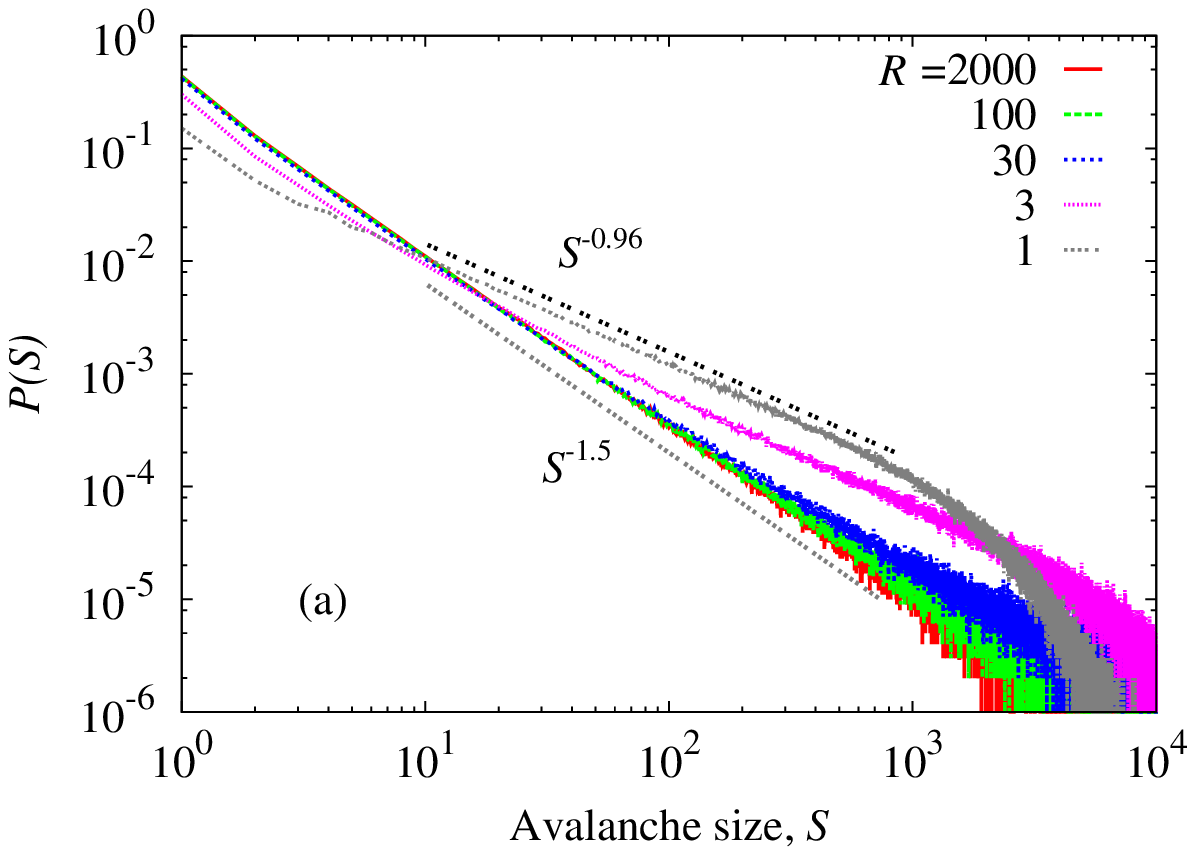}
\includegraphics[height=6cm]{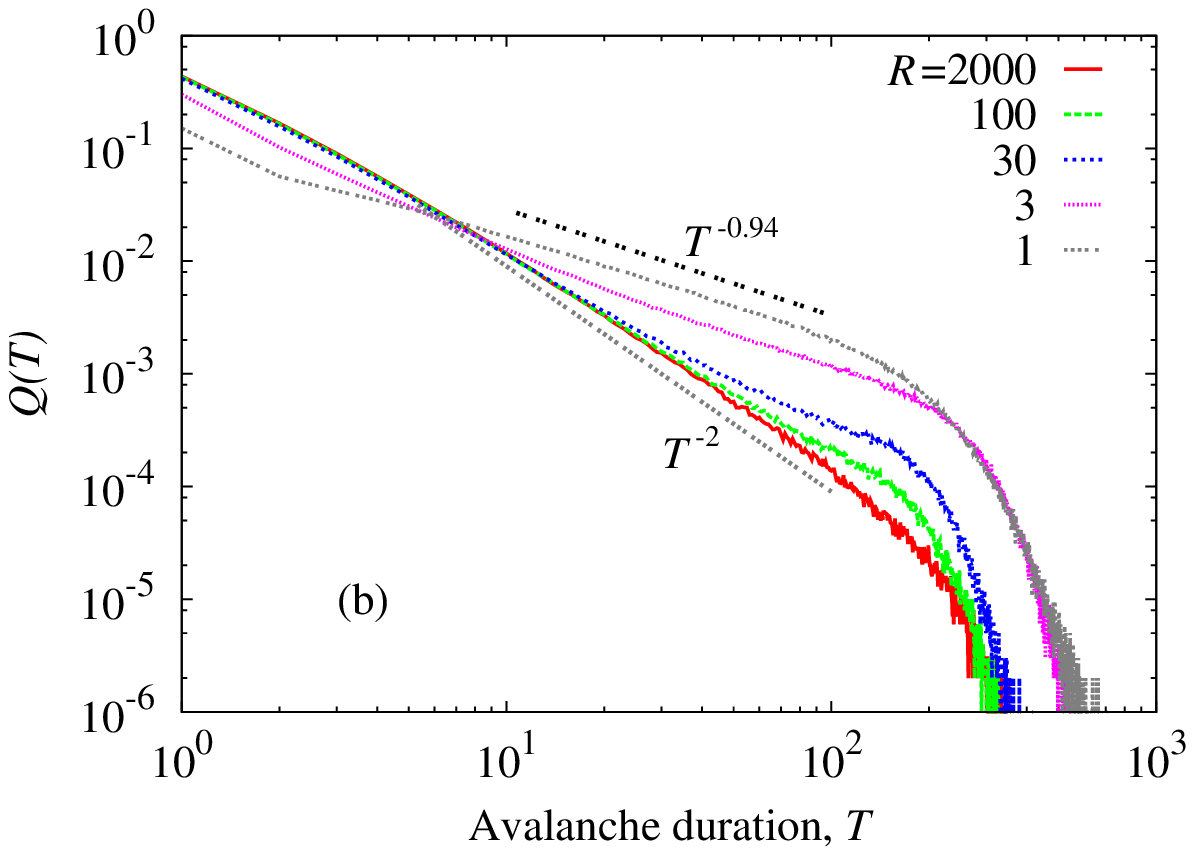}
   \caption{The size (a) and duration (b) distributions of avalanches are shown for different values of $R$
and a system size $L=10000$.
A crossover in the exponent values for size (duration) is seen from a mean-field value of 1.5 (2) to a local
value of 0.96 (0.94). The latter scaling exponents are best-fit values to the numerical results at $R=1$. As discussed in the text, smaller avalanches do not see 
the load sharing range $R$, hence follow mean-field statistics. }
\label{ava_size}
\end{figure}

The total number of fibers breaking between two successive loading steps, denoted by $S$,
measures the size of  an avalanche. The number of load redistribution steps before the avalanche stops is the duration of the avalanche $T$. 
 In Fig. \ref{ava_size} the size distribution, $P(S)$,  and duration distribution, $Q(T)$, of the avalanches are 
plotted for different values of $R$. 
Both distributions show a crossover in the exponent value of a power-law fit, beyond a given size (or duration), 
and the crossover size (duration) increases with increasing $R$. 
While a precise crossover scale will be obtained later (sec.\ref{sec:crossover}), from these plots it is intuitively clear that for 
a given $R$ small avalanches will not
see the scale of load redistribution and will give mean-field statistics, since the load distribution within that range is uniform.
Only the larger avalanches, involving several redistribution steps, 
 will feel the effect of the load redistribution range and hence deviate from mean-field behavior. 

\subsubsection{Roughness exponent of the front:} 
\label{sec:roughness}
The  response of the system in the vicinity of the depinning critical point can be studied
from its dynamics, as done above, and also from its static characteristics, for example in the roughness of the
propagating front. 
 The self-similar nature of the front, at any given moment, in the case of fracture 
is well studied (see Ref. \cite{bonamy11} for a review). In order to demonstrate the roughness properties of our model, we measure
the amplitude of the height fluctuations
\begin{eqnarray}
C(r)=\langle(h_{i+r}-h_i)^2\rangle^{1/2}\sim r^{\zeta_r},
\label{rough_measure}
\end{eqnarray}  
where the angular bracket is the average over space and $\zeta_r$ is the roughness or Hurst exponent. This exponent can also be estimated in Fourier space \cite{tanguy98}, via the decay of the power spectrum of $h$,
\begin{equation}
\langle |\tilde{h(k)}|^2\rangle \sim k^{-1-2\zeta_k},
\label{roughness_measure2}
\end{equation} 
where $\zeta_k$ is another measure for the roughness exponent.
Both the auto-correlation and power-spectral methods attempt to describe the same thing, namely the scaling exponent of a 
self-affine curve. Usually they are equivalent, except in certain limits,
such as the case of very small (or even negative) roughness exponents \cite{hansen01, yang95}.
Here we will focus on the more common, real-space measurement. However, in sec. 3.3,
when small/negative exponents are expected, we will compare the two methods.

\begin{figure}[tbh]
\centering 
\includegraphics[height=6cm]{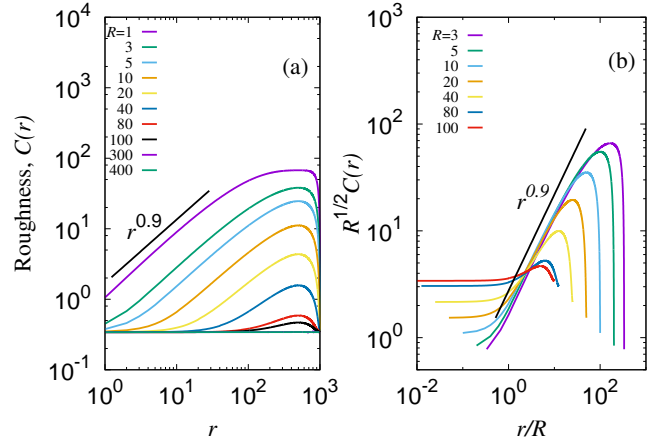}
   \caption{(a) The self-similar scaling of the roughness of the front is shown, following the measure
given in Eq. (\ref{rough_measure}). The height fluctuations in the mean-field limit ($r/R\to 0$) show no variation with 
 $r$, since
there is no notion of distance below the length scale $R$. (b) When distance is
scaled by $R$, all crossover points collapse to the same value, as does the scaling
of the roughness both above and below the point $r=R$. The system size is $L=1000$.}
\label{Roughness_R}
\end{figure}

Following the measure defined in Eq. (\ref{rough_measure}),
Figure \ref{Roughness_R}(a) depicts the behavior of the roughness as the interaction range $R$ is varied. 
As in the case of avalanche statistics, the scaling functions also show
a crossover in critical exponent, depending on the value of $R$. Clearly,  as long as $r<R$, $C(r)$ 
should be independent of
$r$, since there is no notion of spatial distance within that range. In other words, over 
distances shorter than the interaction length $R$, the system should appear as mean-field-like. 
It is only when we look for correlations over  large enough length scales (i.e. bigger than $R$),
that anything else would be expected. In particular, for $r>R$, we find a non-trivial scaling exponent, with  
$\zeta_r\simeq 0.9$.
This point is
further clarified in Fig. \ref{Roughness_R}(b), where the distance $r$ is normalized by $R$. All curves then  show the same
crossover in roughness near the point $r\approx R$. 

\subsubsection{Crossover scale}
\label{sec:crossover}
For a more precise measure of the crossover scale, we look at the long time average of the load per fiber 
(which is also the critical stress $\sigma_c$)
on the system for different values of $R$. In the mean-field limit, the stress is
expected to saturate to $1/3$,  as was
the case in Ref. \cite{biswas13}. We therefore look at the quantity $\Delta\sigma_c=\sigma_c(R,L)-1/3$, which will
tend to zero as either the interaction range $R$ or the system size $L$ is increased. 
For a given system size, however, finite size effects
will make $\Delta \sigma_c$ saturate at a small but finite value. A crossover scale, consistent with our usage in
sec.\ref{sec:roughness}, can be defined 
from the point where $\Delta \sigma_c$ starts becoming independent of $R$. Quantities behaving
independently of $R$ are the signature of the mean-field limit.

\begin{figure}[tbh]
\centering 
\includegraphics[height=6cm]{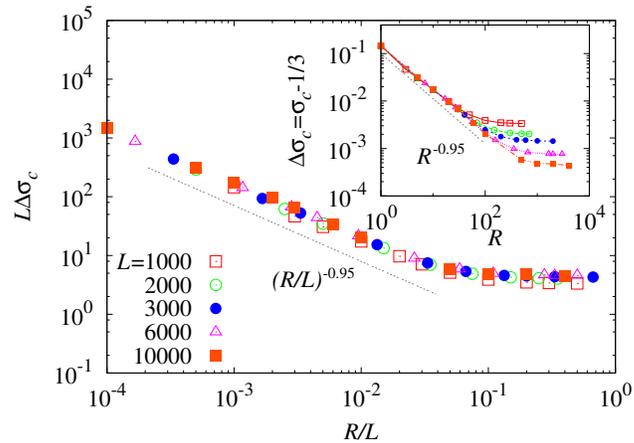}
   \caption{The inset shows how the critical load approaches the mean-field limit as $R$ 
increases. Beyond some crossover value $R_c$, the quantity  $\Delta \sigma_c$ becomes independent of $R$, signaling the change
from local to mean-field behavior.
The main panel  shows how all data collapses when scaled by system size,
 confirming that $R_c\sim L$. In other words,  when the interaction range $R$ is larger than roughly 10\%
of the system size, the system behaves like the mean-field case.}
\label{crpt_R}
\end{figure}
In Fig. \ref{crpt_R} the dependence of  $\Delta\sigma_c$ on the redistribution range $R$ for different system sizes
is plotted. 
The inset shows how the crossover behavior depends on the system size, and also demonstrates how
$\Delta\sigma_c$ decreases to zero as the system size grows. Data collapse is obtained
by scaling the distance with $L$. This implies that $R_c\sim L$, or that a significant fraction of the
system must be covered by the load sharing range in order for the system to show mean-field like behavior.

\subsection{Intermittent dynamics with dissipative scale-free interactions}
Finally, we study the effects of varying the load redistribution process.  Now, following a failure at site
$i$, the load received at site $j$ will be proportional to $\rho=\frac{1}{|i-j|^{\gamma}}$.
As one example of such a response,
  linear-elastic theory predicts an inverse square ($\gamma=2$)
\cite{sch95, ram97} load redistribution for an infinite sample. However, the redistribution range and scaling behavior can be modified due to effects such as a 
finite sample \cite{hutch, bares14}, the width of the sample \cite{xia00}, any correlation in the local disorder \cite{peng91}, 
and so on. In such situations, a scale 
free load redistribution is
closer to what one can expect, instead of the finite range interaction presented in the preceding sections.

 The case $\gamma=2$, which corresponds to a classic Inglis crack,
was studied in Ref. \cite{pontuale13}. But here we show that the $\gamma=2$ case lies in the middle of a continuously varying 
`universality' that interpolates between local and global scenarios. Hence a small change in $\gamma$ can lead to a significant change
in macroscopic observables, such as activity statistics and roughness exponents.

\begin{figure}[tbh]
\centering 
\includegraphics[width=9cm]{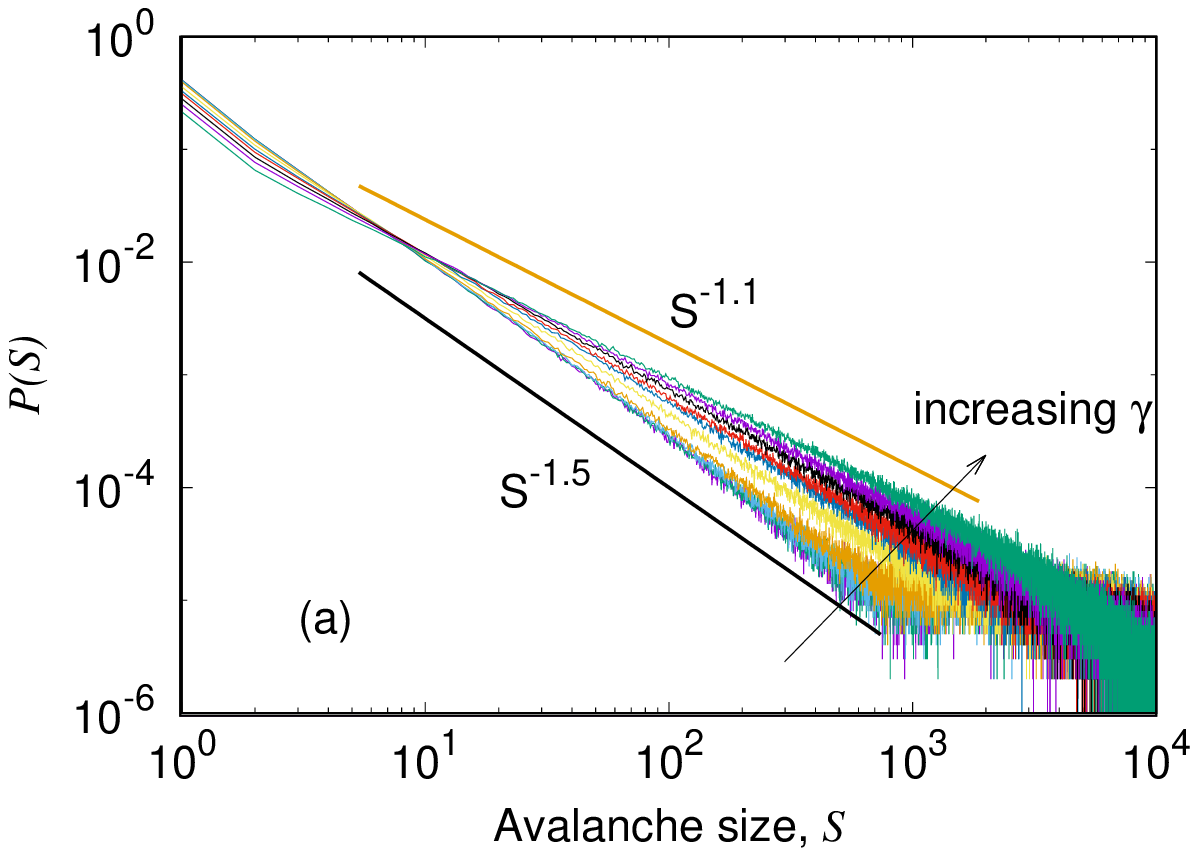}
\includegraphics[width=9.5cm]{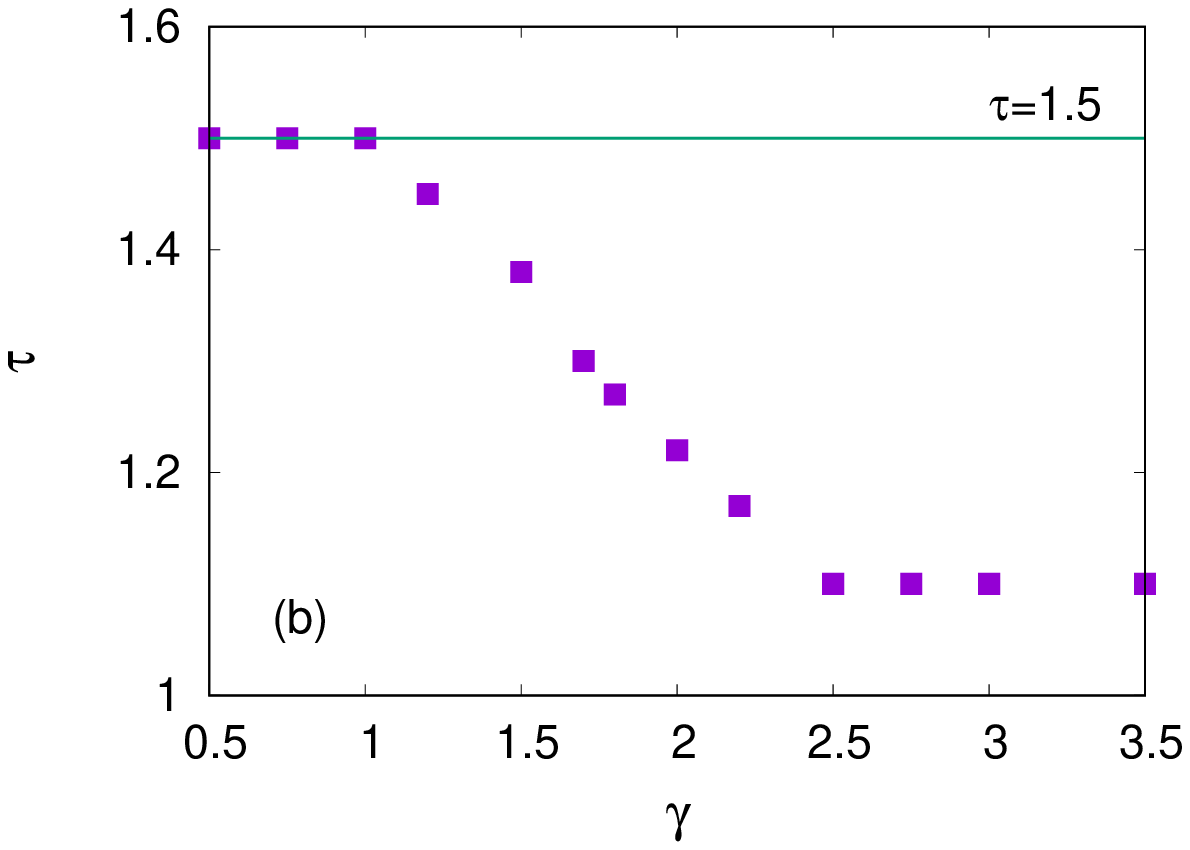}
   \caption{(a) The avalanche size distributions $P(S)$ are plotted for different values of $\gamma$. 
They have the scale invariant form $P(S)\sim S^{-\tau}$. The exponent $\tau$ varies with $\gamma$ as
 shown in  panel (b). As mentioned in the text, the departure from the mean-field 
exponent 1.5 is seen for $\gamma>\gamma_c=1$. The system size is $L=10000$.}
\label{ava_expo}
\end{figure}
Before summarizing the numerical results, let us briefly look into how the crossover value of
the exponent $\gamma$ may be predicted from the linear scaling of $R_c$ with system size obtained in the previous section.
 Even for
a `scale-free' distribution there are still two obvious scales: the lower cut-off scale of the interaction range is the 
lattice spacing and upper cut-off is the system size. Hence it is possible to define an effective
range $R_{eff}=\int \limits_1^{L}r\rho(r)dr$ with $\rho(r)$ being the (normalized) power-law redistribution of load. 
It can be shown \cite{biswas15} that $R_{eff}=\frac{1-\gamma}{2-\gamma}\frac{L^{2-\gamma}-1}{L^{1-\gamma}-1}$.
For $1<\gamma<2$, $R_{eff}\sim L^{2-\gamma}$, which must match with $R_c\sim L$ at the crossover point.
This gives $2-\gamma_c=1$ i.e. $\gamma_c=1$. So for any  value of $\gamma$ larger than 1, the average redistribution range of the load will be smaller than the system size, and thus mean-field-like behavior can no longer be expected. 
 However, as we shall see, this deviation from the mean-field response does not necessarily imply local load sharing behavior for other measures, such as the avalanche and roughness statistics.

Note that $\gamma_c=1$ is also an intuitive result from the fact that for $\gamma\le 1$ the effective range diverges for 
an infinite system. But that does not exclude the possibility of a mean-field response where $\gamma_c>1$, which is indeed the case for
catastrophic breakdown \cite{biswas15}. Therefore, the above result is a joint consequence of the linear scaling 
$R_c\sim L$ and the divergence of the effective range for $\gamma\le 1$.

\begin{figure}[tbh]
\centering 
\includegraphics[height=6cm]{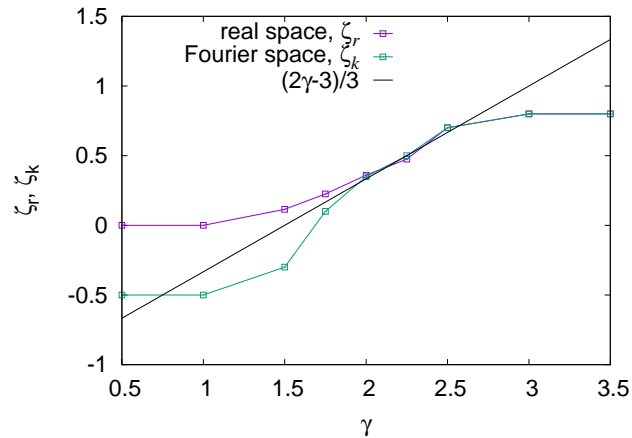}
   \caption{The roughness of the front produced for scale free load redistribution is measured by analyzing the auto-correlation (Eq. \ref{rough_measure}) as well as the power spectrum (Eq. \ref{roughness_measure2}).  The continuous variation of the roughness exponents $\zeta_r$ and $\zeta_k$ with  $\gamma$
is obtained. Near $\gamma=2$ they match with the theoretical prediction \cite{narayan93, tanguy98}, shown by the straight line, of a fluctuating elastic line. For small values of $\gamma$, the negative roughness exponent values are obtained for Fourier space measurements, which can not be seen in the real-space measurements. The system size is $L=1000$.}
\label{roughness_expo}
\end{figure}
In Fig. \ref{ava_expo}(a), the avalanche size distributions are plotted for different $\gamma$ values. They show a continuous 
variation in the exponent  $\tau$ of the size distribution, $P(S)\sim S^{-\tau}$, as detailed in Fig. \ref{ava_expo} (b).  
As expected, the departure from the mean-field value ($\tau=1.5$) starts from just above $\gamma_c=1$.

To measure the roughness exponent, we here use both the real-space (autocorrelation)
method of Eq. (\ref{rough_measure}) and the Fourier-space (power-spectral) method
of Eq. (\ref{roughness_measure2}). The latter technique is more general, and is
capable of measuring negative roughness exponents, such as are expected for example in the case of
white noise (where $\zeta_k=-0.5$) \cite{batrouni02}. Power law fits are used to find the respective exponents
$\zeta_k$ and $\zeta_r$ for each $\gamma$, and these roughness exponents are shown in Fig. \ref{roughness_expo}.
 For the similar situation of a  fluctuating elastic line, both
theory \cite{narayan93} and simulations \cite{tanguy98} predict a relation $\zeta_k=(2\gamma-3)/3$, which is also plotted
in the graph. Around $\gamma=2$ it follows our simulation points well. Away from 
this intermediate case, there is departure from the elastic line simulations, in particular for high $\gamma$ values. 
This further underlines the difference between the macroscopic modeling approach of the elastic line
and the microscopic discrete model studied here. They differ qualitatively as the interaction range becomes
localized (high $\gamma$), while keeping the correspondence near the linear-elastic regime at $\gamma=2$.

\section{Discussion and conclusion}
The common features that arise in various physical systems such as
domain walls in magnetic materials, vortex lines in superconductors, charge density waves, contact line
dynamics in wetting and also in mode-I fracture front, is that an ``elastic" line or front is driven by an external parameter
(a magnetic field, mechanical force, etc.) through a medium with many obstacles (impurities, variations in
breaking strength, etc.). This results in intermittent dynamics of the moving front. The classification of
the properties of such a front depends on its elastic nature, or in other words, the range upto which
the front is affected following a local perturbation. 

Here we have reported the critical properties of such a front through a simple, discrete fiber bundle model approach.
The different strengths of the fibers represent the disorder in the material. 
While this is a generic model for interface propagation in the cases mentioned above, we focused on
mode-I fracture front propagation in this study.
Assuming the applied load to be localized along a line, we can track the activity of that line. We aimed at determining the influence of a variable range of interaction, characterizing 
the load redistribution range following
the failure of a fiber along the interface, on macroscopic static and dynamic properties of the interface, such as
roughness and avalanche statistics. We found such macroscopic observables to be very sensitive to 
the range of interaction. In particular, we 
could identify a crossover from local, nearest-neighbor-like behavior to global, mean-field-like behavior by varying the
range of load redistribution. We showed how this could be done directly, using a flat distribution over a finite neighborhood, or   indirectly  by varying the exponent $\gamma$ in the case of a power-law load redistribution.
The crossover range $R_c$ scales linearly with the system size in the case of a finite range redistribution. This situation
corresponds to
$\gamma_c=1$, for which the behavior deviates from the mean-field in the case of power-law load 
redistribution. For a power-law load redistribution, a continuously varying set of exponents are obtained for the avalanche
size statistics and the roughness of the front. Interestingly, the $\gamma=2$ case lies in the middle of such a continuous
variation, implying that any change in $\gamma$ will be reflected in the roughness and avalanche size exponent values.

An obvious point of comparison for these results is with the elastic interface model \cite{rice}. This  model 
is formulated from 
an opposite approach to ours, in the sense that it is a macroscopic model. Models of this type consider  the
elastic interactions of the interface line from the perspective of linear-elastic fracture mechanics, where the  
disorder of the material is treated as 
quenched noise, as opposed to a microscopic modeling of the
disordered solid.  The elastic interaction can fall off with distance $r$ from a local perturbation as $1/r^{\gamma}$. 
Similar characterization
by varying the parameter $\gamma$ leads to a continuous variation in the roughness exponent value \cite{tanguy98},
 and also crossover behavior \cite{chen15}.
We find the mean-field limit of these two approaches are the same; both in terms of roughness of the interface,
which lacks spatial correlation at this point, and also in terms of the avalanche size distribution. Particularly, the roughness exponent becomes $-0.5$ and avalanche size distribution exponent becomes $1.5$ for
both of these models. Also, for $\gamma=2$, the roughness exponents are close to each other and this also compares well with the 
result in \cite{gjerden13}, where the elastic line limit was observed in the large scale limit.
 But apart from the mean-field limit, or dimensions higher than 4 \cite{ledoussal}, they are both quantitatively
(e.g. Fig. \ref{roughness_expo} for higher $\gamma$ values) and qualitatively different
 (e.g. for fiber bundle the depinning transition is 
discontinuous for
the nearest neighbor interaction) from the fiber bundle model of an interface. 

Many experiments have been done on fracture front propagation in disordered solids in the last couple of decades \cite{book2, book3}.
Among other things, they concern themselves with the in and out-of-plane roughness properties of the broken surfaces, measured both 
along and perpendicular to the direction of the crack front \cite{bonamy11}.
The situation is somewhat simplified when out of plane roughness is suppressed and the crack propagates 
along an easy plane only, resembling the in-plane motion of the front studied here.  
Particularly, in Ref. \cite{pdms1} 
two Plexiglas plates were sandblasted, joined together and then slowly forced open from one edge such that a
mode-I crack opening took place. The propagating front was viewed
by the loss of transparency of the joined plates. The avalanche dynamics and roughness properties of such fronts
are well studied \cite{maloy01, maloy06, san10, chopin15}. Apart from showing a crossover in response from 
small-to-large length scales, the
roughness exponent measurements also show differences in these two limiting regions.  The  
avalanche exponent can be significantly lower (1.1) than 
predicted (1.28) by the elastic interface
model with inverse square elastic strength \cite{bares14}, but also there exists a large spread in the measured values
of such exponents in different experiments \cite{bonamy09}. Such a spread is not unique to the case of fracture front
propagation, but is also seen in the cases of domain wall dynamics in magnetic systems \cite{domain1, domain2, domain3},
and fluid invasion fronts in porous media \cite{he92}. 
Although several mechanisms were proposed to explain these differences, such as driving rate, correlation in
disorder, microscale coalescence and so on, no consensus has yet been reached.
Our simulation shows how one can relate such a change  in exponent values with the variation of an effective interaction range, 
which in turn can be due to one or more of the reasons mentioned above. 

In conclusion, the interface propagation in the fiber bundle model shows a crossover behavior from local to
mean-field statistics as the range of the interaction is increased. The crossover length scales linearly 
with the system size. This implies that for power-law load sharing for exponent values greater than one, the behavior
will be different from the mean-field. This simple prediction is also confirmed numerically. A continuous variation in the 
avalanche size and roughness scaling exponent values were seen with different powers of the scale-free load
sharing, making it a very relevant parameter to keep in mind while considering the finite scales of interaction that might arise
in experimental situations.

\appendix
\section{Depinning transition exponents in the mean field limit}
When $R\sim L$, the depinning transition seen in our model is the mean-field type. 
The growth of the order parameter, here the activity rate, above the critical load behaves
as $A\sim (\sigma-\sigma_c)^{\theta}$. Assuming the scaling form 
\begin{equation}
A(t)\sim t^{-\alpha}F\left(t^{1/\nu}|\sigma-\sigma_c|,t^{z}/L\right),
\end{equation}
we can estimate the scaling exponents $\alpha$, $\nu$ and $z$ and also verify the scaling relation 
$\alpha=\theta/\nu$, in this limit.

\begin{figure}[tbh]
\centering 
\includegraphics[height=6cm]{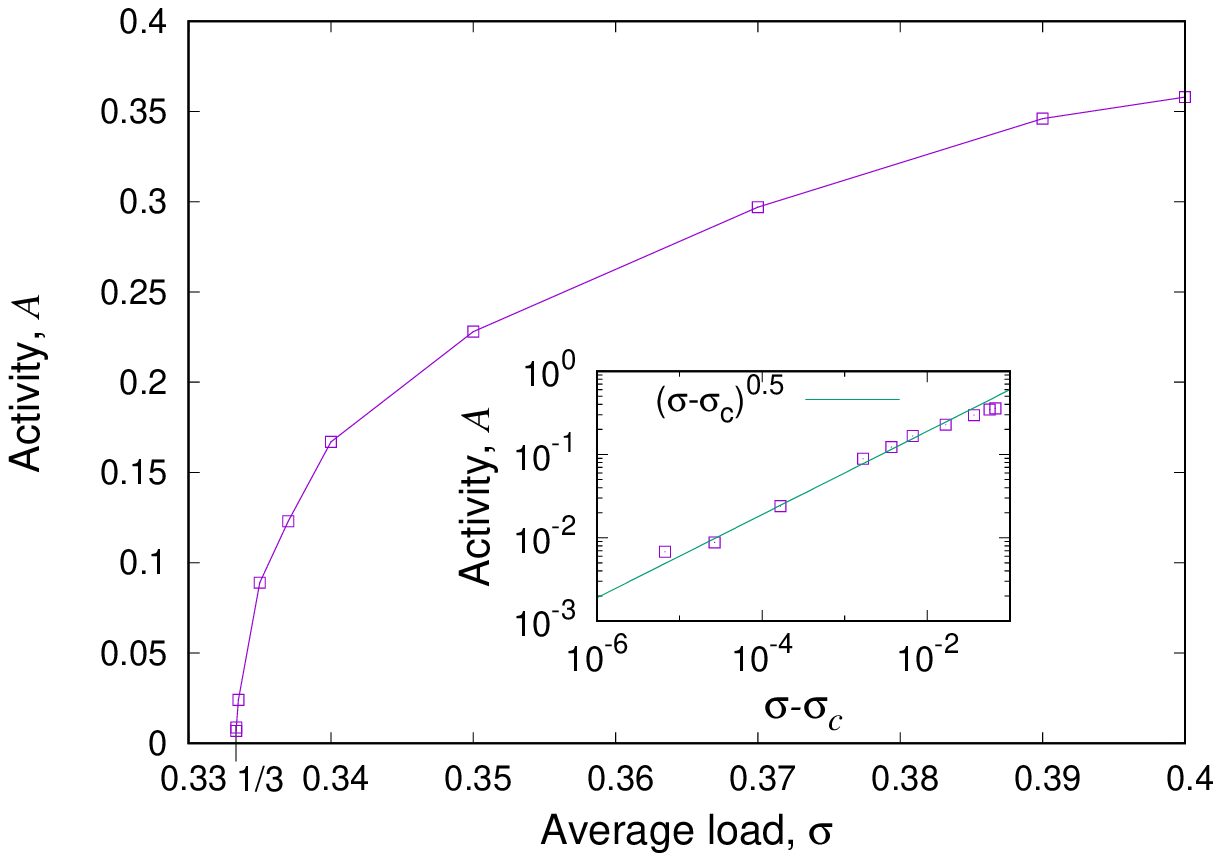}
   \caption{The growth of the order parameter $A$ beyond the critical load $\sigma_c$ is shown. The inset shows the power-law fitting
with an order parameter exponent value $\theta=1/2$. The system size is $L=700~000$.}
\label{mf_depin_3}
\end{figure}
\begin{figure}[tbh]
\centering 
\includegraphics[height=6.5cm]{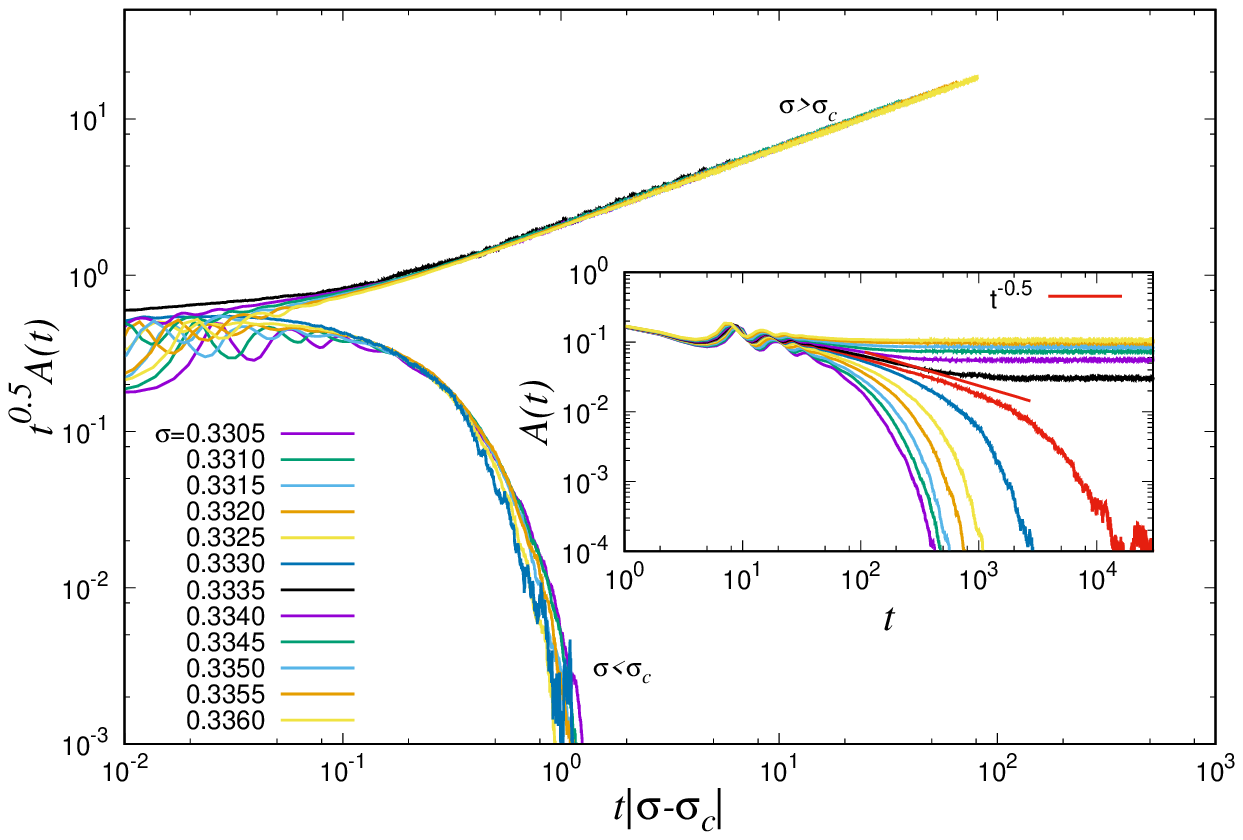}
   \caption{The super-critical and sub-critical behavior of the order parameter with time are shown. At the critical point
a power-law decay gives the exponent $\alpha\approx 0.5$ and the off-critical collapse shows gives the scaling
exponent $\nu=1.0$. The system size is $L=100~000$.}
\label{mf_depin_1}
\end{figure}
In Fig. \ref{mf_depin_3} we plot the saturation activity rates at different $\sigma$ values above $\sigma_c=1/3$
for the mean-field case of $R=L$. 
The inset shows the power-law variation $(\sigma-\sigma_c)^{1/2}$, giving $\theta=1/2$, which
matches with the prediction for the mean-field elastic interface \cite{van}. At the
critical point the order parameter decays as $A(t)\sim t^{-\alpha}$, where $t$ is the time measured by the number of 
load redistribution steps. This is seen in Fig. \ref{mf_depin_1}, 
with $\alpha\approx 0.5$. Also, by plotting $A(t)t^{\alpha}$ with $t|\sigma_c-\sigma|^{\nu}$
and by tuning $\nu$, we get a data collapse near $\nu\approx 1.0$ (see Fig. \ref{mf_depin_1}). This agrees with 
the scaling relation $\alpha=\beta/\nu$. Finally, for the finite size scaling
at the critical point, the cut-off time scale $\tau$ scales with system size as  $\tau\sim L^{1/z}$ with $z\approx 2.0$.
This set of exponents defines the mean-field depinning transition for fiber bundle interfaces, and these results show
 that our model reaches this well-defined response in the appropriate limit.
 
\section*{Acknowledgements}
SB thanks the Alexander von Humboldt foundation for funding.

\end{document}